\documentclass[sn-basic]{sn-jnl}% Basic Springer Nature Reference Style/Chemistry Reference Style
\usepackage{graphicx}%
\usepackage{multirow}%
\usepackage{amsmath,amssymb,amsfonts}%
\usepackage{amsthm}%
\usepackage{mathrsfs}%
\usepackage[title]{appendix}%
\usepackage{xcolor}%
\usepackage{textcomp}%
\usepackage{manyfoot}%
\usepackage{booktabs}%
\usepackage{algorithm}%
\usepackage{algorithmicx}%
\usepackage{algpseudocode}%
\usepackage{listings}%

\usepackage{bm}
\renewcommand{\vec}[1]{{\bm{#1}}}

\begin{document}

\title[NMF as GCM]{Applying Non-negative Matrix Factorization with Covariates to the Longitudinal Data as Growth Curve Model}

\author[1]{\fnm{Kenichi} \sur{Satoh}}
% https://orcid.org/0000-0003-4436-9347
\email{kenichi-satoh@biwako.shiga-u.ac.jp}
\affil[1]{\orgdiv{Faculty of Data Science}, \orgname{Shiga University}, \orgaddress{\street{Banba 1-1-1}, \city{Hikone}, \postcode{522-8522}, \state{Shiga}, \country{Japan}}}


\abstract{Using Non-negative Matrix Factorization (NMF), an observed matrix is approximated by a basis matrix times a coefficient matrix. When each individual's coefficient vector is explained by covariates, the coefficient matrix factorizes into a parameter matrix and a covariate matrix---a tri-factorization whose mean structure coincides with that of the Growth Curve Model (GCM) for longitudinal data. This correspondence has been noted but not examined. We make three contributions. First, we compare NMF with covariates and the GCM: the basis is prescribed in the GCM but optimized in NMF, and the NMF-optimized basis can be used within the GCM and may improve its fit, the two agreeing when covariate effects are non-negative. Second, the main contribution, we develop statistical inference for the parameter matrix linking covariates to basis components: conditional on the optimized basis we provide standard errors, Wald-type tests, and one-sided confidence intervals, with a simulation study confirming good calibration for covariate-effect contrasts. Third, we compare NMF with principal component analysis (PCA) and functional PCA (FPCA): its non-negative coefficients are membership probabilities giving a soft clustering directly, whereas signed PCA/FPCA scores require a downstream classifier. Illustrations use growth data and a kernel-based varying-coefficient model.
}


\maketitle

%-----------------------------------------------------------
\section{Introduction}\label{sec1}
%-----------------------------------------------------------

Non-negative Matrix Factorization (NMF) was proposed by \citet{lee1999} to approximate an observed matrix consisting of non-negative elements as the product of a basis and coefficient matrices. It is often regarded as a dimension-reduction method because it represents multidimensional data with lower-dimensional bases; however, it can also be regarded as a regression model with latent variables, in which the coefficients act as latent variables (scores) and the unknown basis is estimated jointly with them from the data.
Moreover, because the coefficients are non-negative and can be interpreted as proportions or probabilities, they are used for soft clustering \citep{ding2005}.
NMF is a versatile analytical method for both supervised and unsupervised machine learning.
NMF has been applied in various fields, such as image data object recognition by \citet{lee1999}, polyphonic music transcription by \citet{smaragdis2003}, and text data by \citet{berry2009} and \citet{mifrah2020}.

Optimizing all entries of the basis and coefficient matrices jointly is a non-convex problem with no closed-form solution, so directly searching its many parameters is impractical. However, the multiplicative update of \citet{lee2000} decreases the objective function at every step from any non-negative starting value. Because each step is guaranteed not to increase the objective and requires no matrix inversion or step-size tuning, the optimization is numerically stable even when the dimension of the observed vector is large, which makes the method practical for large data sets. Moreover, the analyst can readily add constraints as needed.

Ordinary NMF, however, uses only the observation matrix and no covariates, so it cannot predict from background or risk factors. On the other hand, the Growth Curve Model (GCM) of \citet{potthoff1964}, long used for longitudinal data, expresses the mean as the product of three matrices---a basis (design) matrix, a parameter matrix, and a covariate matrix. Modelling the NMF coefficient matrix by a known covariate matrix, as the product of a parameter matrix and that covariate matrix, gives the observation matrix exactly this three-matrix form, so that NMF acquires the GCM mean structure and can thus incorporate covariates. The present formulation was in fact reached from the GCM side rather than from NMF, and through a sequence of studies these two traditions converge. First, the GCM, originally developed for longitudinal data, was applied to spatial data with location covariates in \citet{satoh2020jp}. Next, after adopting NMF, the estimated coefficients were regressed on covariates in a separate, second step in \citet{satoh2022jp}. Finally, this coefficient model was folded into a single non-negativity-preserving NMF optimization in \citet{satoh2023}, which observed that the result is formally identical to the GCM mean structure. This is the formulation used here.

The same three-matrix form is also that of the Non-negative Matrix tri-Factorization (tri-NMF) of \citet{ding2006} (see also \citet{gillis2020}), in which all three non-negative matrices are unknown and optimized. The GCM, tri-NMF, and NMF with covariates thus share this form but differ in which matrices are known: in the GCM both the basis (design) matrix and the covariate matrix are prescribed by the analyst and only the parameter matrix is estimated; in tri-NMF all three are optimized; and the present formulation lies between the two, optimizing the basis and the parameter matrices as in NMF while taking the covariate matrix as known as in the GCM. Because the covariate matrix is the distinctive ingredient, we call this NMF with covariates. Although tri-NMF is the most general of the three, fixing the covariate matrix to a known one is what broadens applicability---enabling covariates, prediction, and connections to established models---so that NMF with covariates also has a broad range of application (Section~\ref{sec4}).

This study makes three contributions, the second being the main methodological one.
\begin{itemize}
\item \textbf{Comparison with the growth curve model.} The mean structure of NMF with covariates coincides with that of the GCM; although this was noted by \citet{satoh2023}, the two models were not compared. We carry out the comparison and, in particular, show that the basis optimized by NMF can be used within the GCM, where it may improve the fit.
\item \textbf{Statistical inference for the parameter matrix $\Theta$} (the main methodological contribution). Conditional on the basis optimized by NMF, we develop standard errors, Wald-type tests, and confidence intervals for the covariate effects, with a one-sided boundary test for the non-negative parameters, and validate the procedure by a simulation study.
\item \textbf{Comparison with PCA and FPCA.} Treating NMF as a dimension-reduction and clustering method for longitudinal data, we compare it with principal component analysis and functional principal component analysis.
\end{itemize}

Section \ref{sec2} introduces NMF with covariates. Section \ref{sec3} presents the update formulas for optimization in the tri-NMF framework. In Section \ref{sec4}, we describe its relationship with GCM.
In Section \ref{sec:inference}, we develop statistical inference for the parameter matrix $\Theta$.
Section \ref{sec5} presents applications to real data, including the comparison with PCA and FPCA and a Gaussian-kernel varying-coefficient model; Section \ref{sec:sim} validates the proposed inference through a simulation study calibrated to one of these examples; and Section \ref{sec6} gives the discussion and conclusion.

%-----------------------------------------------------------
\section{Non-negative Matrix Factorization with covariates}\label{sec2}
%-----------------------------------------------------------

We build the model from the individual up. For individual $n$, the non-negative observation vector $\vec{y}_n=(y_{1,n},\dots,y_{P,n})'$ of $P$ measured variables---the response---is approximated by a basis matrix $X=(\vec{x}_1,\dots,\vec{x}_Q)=(x_{p,q})_{P \times Q}$ that is common to all individuals together with an individual-specific non-negative coefficient vector $\vec{b}_n=(b_{1,n},\dots,b_{Q,n})'$, as $\vec{y}_n \approx X\vec{b}_n$; that is, each individual's measurements are a non-negative combination of the $Q$ common basis patterns weighted by $\vec{b}_n$. Collecting the $N$ individuals as the \emph{columns} of $Y=(\vec{y}_1,\dots,\vec{y}_N)=(y_{p,n})_{P\times N}$ and $B=(\vec{b}_1,\dots,\vec{b}_N)=(b_{q,n})_{Q \times N}$ places these per-individual models side by side and gives the non-negative matrix factorization (NMF) of \citet{lee1999,lee2000},
\begin{equation}
\mathop{Y}_{P \times N} \approx \mathop{X}_{P \times Q}\mathop{B}_{Q \times N}.
\label{eq01}
\end{equation}
The number of bases $Q$ is chosen by the analyst. Since a non-negative matrix has non-negative rank at most $\min(P,N)$, taking $Q>\min(P,N)$ only adds redundant basis vectors without improving the approximation, so for approximating $Y$ it suffices to take $Q\le\min(P,N)$ (and, for dimension reduction, $Q\ll\min(P,N)$).
A column of $Y$ is therefore one individual and a row is one variable, so $Y$ is $P\times N$ (variables by individuals)---the transpose of the data matrix used in much of statistics, in which individuals are rows. Arranging individuals as columns is precisely what lines the per-individual models $\vec{y}_n\approx X\vec{b}_n$ up as the columns of (\ref{eq01}).

The factorization (\ref{eq01}) is not unique: for any positive diagonal matrix $D$, replacing $X$ by $XD$ and $B$ by $D^{-1}B$ leaves the product unchanged. We therefore normalize each column of $X$ to sum to one. This normalization is not required for the optimization---it may be omitted---but it removes the scale indeterminacy and improves both the interpretability and the numerical stability of the solution. Each column of $X$ is then a probability vector over the $P$ variables, and the share $b_{q,n}/\sum_{q'} b_{q',n}$ of individual $n$ on basis $q$ serves as a membership probability, so the non-negative coefficients provide a soft clustering \citep{ding2005}.

Next, we explain the coefficient matrix $B$ using covariates. In other words, we use the known covariate matrix $A=(\vec{a}_1,\dots,\vec{a}_N)=(a_{r,n})_{R\times N}$ and the parameter matrix $\Theta=(\theta_{q,r})_{Q \times R}$, both having non-negative elements, to write
\begin{equation}
\mathop{B}_{Q \times N}=\mathop{\Theta}_{Q \times R} \mathop{A}_{R\times N}
\label{eq02}
\end{equation}
Thus, $Y\approx XB=X\Theta A$. Additionally, for individual $n$, we have
\begin{equation}
\vec{y}_n \approx X\vec{b}_n=X\Theta \vec{a}_n,
\qquad n=1,\ldots,N,
\label{eq03}
\end{equation}
Table~\ref{tab:notation} summarizes the matrices and operators used throughout.

\begin{table}[h]
\caption{Notation. Matrices are written in upper case and column vectors in bold lower case; a prime denotes transpose.}\label{tab:notation}
\centering
\begin{tabular}{ll}
\toprule
Symbol & Meaning \\
\midrule
$Y=(y_{p,n})_{P\times N}$ & observation matrix: $P$ variables (rows) $\times$ $N$ individuals (columns) \\
$X=(x_{p,q})_{P\times Q}$ & basis matrix ($Q$ basis components) \\
$B=(b_{q,n})_{Q\times N}$ & coefficient matrix \\
$\Theta=(\theta_{q,r})_{Q\times R}$ & parameter matrix \\
$A=(a_{r,n})_{R\times N}$ & covariate matrix \\
$\vec{y}_n,\ \vec{b}_n,\ \vec{a}_n$ & the $n$th columns of $Y$, $B$, $A$ (individual $n$) \\
$M'$ & transpose of a matrix $M$ \\
$\mathrm{tr}(M)$ & trace of a square matrix $M$ \\
$\|M\|_F$ & Frobenius norm, $\|M\|_F^2=\mathrm{tr}(M'M)$ \\
$\odot,\ \oslash$ & element-wise (Hadamard) product and division; $(M\odot N)_{ij}=M_{ij}N_{ij}$ \\
$\otimes$ & Kronecker product \\
$\mathrm{diag}(\vec{v})$ & diagonal matrix with diagonal entries $\vec{v}$ \\
$\vec{1}$ & vector (or matrix) of ones \\
$M\ge0$ & every entry of $M$ is non-negative \\
\bottomrule
\end{tabular}
\end{table}

This approximation model, expressed as the product of three matrices, is included in the tri-NMF, proposed by \citet{ding2006}. Note that in tri-NMF, the three matrices are generally unknown; however, the covariate matrix in this study is known.
Thus, for a new covariate $\vec{a}$, we obtain the coefficient vector $ \vec{b}=\Theta \vec{a}$ and predict the observation value using $\hat{\vec{y}}=X\vec{b}$.

This approximation has the same form as the mean structure of the GCM; see \citet{von1991} for the GCM and Section~\ref{sec4} for a detailed comparison of the two.

It is worth noting the relationship with NMF without covariates. Setting the covariate matrix to the identity, $R=N$ and $A=I_N$, gives $B=\Theta$, so the model reduces to NMF without covariates. Using genuine covariates instead restricts the attainable coefficient vectors and can lower the approximation accuracy relative to the covariate-free case, but in return makes prediction possible. A Gaussian-kernel covariate \citep{satoh2023} mitigates this loss of accuracy; see the bone-density example of Section~\ref{subsec3}.
 
%-----------------------------------------------------------
\section{Optimization of basis matrix and parameter matrix}\label{sec3}
%-----------------------------------------------------------

We introduce update formulas for optimizing the basis matrix $X$ and parameter matrix $\Theta$ when the observation matrix $Y$ and the known covariate matrix $A$ are provided. 

The prediction matrix of $Y$ using NMF is
\begin{equation}
\hat{Y}=(\vec{\hat{y}}_1,\dots,\vec{\hat{y}}_N)=(\hat{y}_{p,n})_{P\times N}=XB=X\Theta A
\label{eq06}
\end{equation}
We measure the discrepancy between $Y$ and $\hat{Y}$ by the squared Euclidean distance, whose minimization is equivalent to maximum likelihood under a Normal model; this is the criterion used throughout the paper. Other divergences are common in NMF and yield analogous multiplicative updates but are not used here: the generalized Kullback--Leibler (KL) divergence, equivalent to maximum likelihood under a Poisson model and suited to count data; the Itakura--Saito pseudo-distance, corresponding to maximum likelihood under multiplicative i.i.d.\ Gamma noise \citep{fevotte2009}; the $\beta$-divergence \citep{cichocki2010}; and the Tweedie distributions \citep{abe2017}.
The update formulas for tri-NMF with a squared Euclidean distance were also presented by \citet{ding2006} and \citet{copar2017} when $X$ and $A$ are orthogonal matrices.
For the numerical computation of NMF without covariates, the package \verb|NMF| of the statistical software R \citep{r2023} by \citet{gaujoux2010} is available.

The update formulas are obtained by differentiating each objective function with respect to $\Theta$ (and $X$) using standard matrix calculus. To preserve non-negativity, the gradient is split into its non-negative parts, $\partial D/\partial\Theta=2(\nabla^{+}-\nabla^{-})$ with $\nabla^{+},\nabla^{-}\ge0$, and $\Theta$ is updated multiplicatively by $\Theta\longleftarrow\Theta\odot(\nabla^{-}\oslash\nabla^{+})$, where $\odot$ and $\oslash$ are the element-wise (Hadamard) product and division; at a fixed point $\nabla^{+}=\nabla^{-}$, so the update leaves $\Theta$ unchanged. The same device is used for $X$.

Explicitly, the squared Euclidean distance is
\begin{equation}
D_{EU}(Y,\hat{Y})=\mbox{tr}\{(Y-\hat{Y})'(Y-\hat{Y})\}=\|Y-X\Theta A\|_F^2 .
\label{eq07}
\end{equation}
Differentiating with respect to $\Theta$ gives $\partial D_{EU}/\partial\Theta=-2X'YA'+2X'X\Theta AA'$, so $\nabla^{-}=X'YA'$ and $\nabla^{+}=X'X\Theta AA'=X'\hat{Y}A'$. The multiplicative update for the parameter matrix is therefore
\begin{equation}
\Theta  \longleftarrow \Theta \odot \{(X'YA')\oslash(X'\hat{Y}A')\}.
\label{eq11}
\end{equation}
With the coefficient matrix $B=\Theta A$ held fixed, $X$ enters only through $\hat{Y}=XB$, so the same argument yields the basis update, which coincides with that of NMF without covariates,
\begin{equation}
X \longleftarrow X \odot \{(YB')\oslash(\hat{Y}B')\}.
\label{eq08}
\end{equation}
 
%-----------------------------------------------------------
\section{Comparison with the Growth Curve Model}\label{sec4}
%-----------------------------------------------------------

The Growth Curve Model (GCM) proposed by \citet{potthoff1964} is as follows:
\begin{equation}
Y \sim N_{P\times N}(X\Theta A, \Sigma\otimes I_N).
\label{eq15}
\end{equation}
The mean structure of GCM and the approximation by NMF with covariates are formally identical and can be described as $X\Theta A$. This correspondence---together with the special cases derived below (the GCM with a fixed basis, and multiple regression)---was noted by \citet{satoh2023}; the present section carries out the comparison in detail, and Section~\ref{sec:inference} develops the statistical inference that the analogy enables.
A comparison of tri-NMF, NMF with covariates, and the GCM is summarized in Table \ref{tab1}.
An unknown matrix is obtained by the multiplicative-update optimization of Section~\ref{sec3} in tri-NMF and NMF with covariates, and by maximum likelihood in the GCM.
In the GCM, $X$ is the within-individual covariate matrix and $A$ the between-individual covariate matrix, both known and given by the analyst, and $\Theta$, the only unknown, has no sign constraint.
In tri-NMF \citep{ding2006}, all three matrices are non-negative, and orthogonality $X'X=I$ and $AA'=I$ is imposed to make the solution unique.
NMF with covariates lies between the two: $X$ and $\Theta$ are unknown while $A$ is known, and uniqueness is instead obtained by normalizing each column of $X$ to sum to one, as explained in Section \ref{sec2}.

\begin{table}[h]
  \caption{Comparison of tri-NMF, NMF with covariates, and GCM}\label{tab1}%
  \centering
  \begin{tabular}{lccc}
  \toprule
            & tri-NMF & NMF with covariates & GCM \\
 \midrule
  Model & $Y \approx X\Theta A$ & $Y \approx X\Theta A$ & $\mbox{E}(Y)=X\Theta A$ \\
  $Y$        & non-negative & non-negative & continuous \\
  $X$ (basis)        & unknown & unknown & known \\
  $\Theta$ (parameter) & unknown & unknown & unknown \\
  $A$ (covariate)        & unknown & known & known \\
  non-negativity & $X,\Theta,A\ge0$ & $X,\Theta,A\ge0$ & none \\
  identifiability & $X'X=I,\ AA'=I$ & $X$ normalized & $X,A$ prescribed \\
  reference & \citet{ding2006} & \citet{satoh2023} & \citet{potthoff1964} \\
\botrule
  \end{tabular}
\end{table}

The basis $X$ optimized by NMF can in turn be used within the GCM, where it may improve the fit, as we illustrate in Section~\ref{subsec1}.

The comparison also reveals an apparent paradox. Of the three, tri-NMF is formally the most general, since all three factors are free; yet that generality serves only to approximate a single observed matrix more flexibly. Fixing $A$ to a known covariate matrix, far from being merely a restriction, is what broadens applicability: external covariates can enter the model, a new unit can be predicted through $\vec{b}=\Theta\vec{a}$ and $\hat{\vec{y}}=X\vec{b}$, and, through the choice of $A$, NMF is connected to a range of established models beyond the growth curve model: for example, vector autoregression for multivariate time series \citep{satoh2026var} and classification through a label matrix \citep{satoh2026lab}. Thus, while tri-NMF is the more general factorization, NMF with covariates also has a broad range of application.

Unlike NMF, the GCM gives each column of $Y$ an unstructured covariance matrix $\Sigma$ with $P(P+1)/2$ parameters.

The maximum likelihood estimators of $\Theta$ and $\Sigma$ on the GCM are obtained by
\begin{equation}
\hat{\Theta}=(X'S^{-1}X)^{-1} X' S^{-1}YA'(AA')^{-1},\qquad
 \hat{\Sigma}=\frac{1}{N}(Y-X\hat{\Theta}A)(Y-X\hat{\Theta}A)',
\label{eq16}
\end{equation}
where $S=Y\{I_N-A'(AA')^{-1}A\}Y'/(N-R)$ and $N-R > 0$.
Thus, the following two points are the greatest advantages of GCM: 1) it does not require repetitive calculations to obtain estimators, and 2) tests and confidence intervals can be constructed based on the distribution of the estimators.

When the basis and covariate matrices of the GCM are non-negative, the non-negative parameter matrix can be optimized using the NMF update formula. From equation (\ref{eq11}), this can be written as
\begin{equation}
\Theta  \longleftarrow \Theta \odot \{(X'YA') \oslash (X'\hat{Y}A')\},
\qquad \hat{Y}  \longleftarrow X\Theta A,
\label{eq17}
\end{equation}
which is iterated to convergence.

Furthermore, the multiple regression model can be described as a special case of GCM. If the dimension of the individual observation data is one, the observation matrix is a row vector $\vec{y}'$ of length $N$ and the basis matrix becomes scalar and can be written as $1$.
Here, the parameter matrix is a row vector $\vec{\theta}'$ of length $R$, which corresponds to the multiple regression model: $\vec{y}' \approx \vec{\theta}'A$
 or $\vec{y} \approx A'\vec{\theta}$.
 Then, the update equation can be written from equation  (\ref{eq17}) as
\begin{equation}
\vec{\theta}  \longleftarrow \vec{\theta} \odot (A\vec{y} \oslash A\hat{\vec{y}}),
\qquad 
\hat{\vec{y}} \longleftarrow   A'\vec{\theta}. 
\label{eq18}
\end{equation}
The resulting regression coefficients are restricted to non-negative values, which makes them less accurate approximations than least-squares estimators; however, they might have the advantage of being easy to interpret. In fact, with $A$ known and $P=1$ (so $X=1$), the only unknown is $\vec{\theta}\ge\vec{0}$ and the problem becomes the convex non-negative least-squares (NNLS) problem \citep{lawson1974,bro1997}; NMF with covariates thus contains the classical NNLS as a special case, which ordinary NMF---whose basis is also unknown---does not.

%-----------------------------------------------------------
\section{Statistical inference for the parameter matrix}\label{sec:inference}
%-----------------------------------------------------------

We now develop inference for the parameter matrix $\Theta$, which quantifies how the covariates affect the basis components. Uncertainty quantification for NMF has been pursued mainly from a Bayesian standpoint, placing priors on the factors and sampling the posterior \citep{schmidt2009,cemgil2009}; these methods quantify the uncertainty of the latent factors but do not provide inference on the covariate effects in NMF with covariates. We therefore develop it here: a classical, frequentist inference for the covariate effects, conditional on the NMF-estimated basis and tied to the growth curve model. NMF itself is a distribution-free approximation method: the multiplicative updates of Section~\ref{sec3} minimize a discrepancy between $Y$ and $X\Theta A$ and make no probabilistic assumption about the data. To attach standard errors and significance levels to the estimated effects, however, a probability model is required. We therefore introduce, \emph{solely for the purpose of inference and without altering the NMF point estimate}, a Gaussian working model. Conditional on the basis matrix $\hat{X}$ optimized by NMF, the squared-error objective is reinterpreted as the negative log-likelihood of
\begin{equation}
Y=\hat{X}\Theta A+\mathcal{E},\qquad \mbox{vec}(\mathcal{E})\sim N(\vec{0},\sigma^2 I_{PN}),
\label{eq:workmodel}
\end{equation}
which is the mean structure of the GCM in (\ref{eq15}) with the unknown design matrix replaced by the optimized basis $\hat{X}$. Vectorizing, $\mbox{vec}(Y)=(A'\otimes \hat{X})\,\mbox{vec}(\Theta)+\mbox{vec}(\mathcal{E})$, which is a linear model in $\mbox{vec}(\Theta)$. The conditional Fisher information is
\begin{equation}
\mathcal{I}(\Theta)=\frac{1}{\sigma^2}\left(AA'\otimes \hat{X}'\hat{X}\right),
\label{eq:fisher}
\end{equation}
so that, treating $\hat{X}$ as fixed,
\begin{equation}
\mbox{vec}(\hat{\Theta})\ \dot{\sim}\ N\!\left(\mbox{vec}(\Theta),\ \sigma^2\left(AA'\otimes \hat{X}'\hat{X}\right)^{-1}\right),
\label{eq:thetadist}
\end{equation}
with $\hat{\sigma}^2=\|Y-\hat{X}\hat{\Theta}A\|_F^2/(PN-QR)$. Standard errors are the square roots of the diagonal of the estimated covariance.

When the information matrix is ill-conditioned, or when one prefers not to rely on the normal approximation near the non-negativity boundary, a \emph{re-estimation} wild (multiplier) bootstrap \citep{wu1986,liu1988} can be used. From the fitted values $\hat{X}\hat{\Theta}A$ and the residuals $\mathcal{E}=Y-\hat{X}\hat{\Theta}A$, each of $B$ replications scales the $n$th residual column by an independent mean-zero, unit-variance multiplier $w_n$, forming $Y^{*}=\hat{X}\hat{\Theta}A+\mathcal{E}\,\mathrm{diag}(\vec{w})$ (negative entries clipped to zero so that $Y^{*}\ge0$), re-estimates the parameter matrix by the multiplicative updates of Section~\ref{sec3} (so that $\hat{\Theta}^{*}\ge0$ automatically), and recomputes the quantity of interest; standard errors and percentile confidence intervals follow from its bootstrap distribution. This differs from a \emph{one-step} (score) bootstrap, which linearizes the estimator by a single Newton step from $\hat{\Theta}$ and so requires the inverse information $\mathcal{I}^{-1}$: by re-solving the optimization instead, the re-estimation form avoids $\mathcal{I}^{-1}$ and remains valid when $\mathcal{I}$ is singular, as for the over-parameterized kernel covariate of Section~\ref{subsec3}.

Because each parameter is non-negative, the significance of an effect $\theta_{q,r}$ is assessed by the one-sided test of the boundary null hypothesis $H_0:\theta_{q,r}=0$ against $H_1:\theta_{q,r}>0$, using $z=\hat{\theta}_{q,r}/\mbox{SE}$. Correspondingly, the confidence interval for a non-negative parameter is one-sided, $[\max(0,\hat{\theta}_{q,r}-z_{\alpha}\,\mbox{SE}),\,\infty)$, rather than a symmetric Wald interval that could extend below zero; a contrast of parameters, being unconstrained in sign, takes a two-sided interval (as for the male-effect band of Section~\ref{subsec3}).

We emphasize that this inference is conditional on the estimated basis $\hat{X}$: it quantifies the uncertainty in $\Theta$ for a given basis and, like the GCM, treats the design matrix as known. Unlike the GCM, however, $\hat{X}$ is data-driven, so the inference is conditional, or selective, in nature \citep{taylor2015}, and the consequences of this are examined by the simulation study of Section~\ref{sec:sim}.

%-----------------------------------------------------------
\section{Examples of data analysis}\label{sec5}
%-----------------------------------------------------------

We analyse the four longitudinal data sets listed in Table~\ref{tab:datasets}.

\begin{table}[h]
\caption{The four longitudinal data sets analysed in this section. $P$ is the number of measurement occasions (rows of $Y$) and $N$ the number of individuals or measurements (columns).}\label{tab:datasets}
\centering
\begin{tabular}{llll}
\toprule
Data set & Response ($P\times N$) & Covariate & Section \\
\midrule
\verb|Orthodont|   & dental distance, 4 ages ($4\times27$)        & sex (intercept $+$ male)            & \ref{subsec1} \\
\verb|ChickWeight| & body weight, 12 days ($12\times45$)          & diet (4 groups)                     & \ref{subsec:chick} \\
\verb|growth|      & stature, 31 ages ($31\times93$)              & sex (intercept $+$ male)            & \ref{subsec:berkeley} \\
\verb|bone|        & relative spinal BMD change ($1\times485$)    & age (Gaussian kernel) $\times$ sex  & \ref{subsec3} \\
\botrule
\end{tabular}
\end{table}

The NMF with covariates used in this analysis is implemented in the R package \verb|nmfkc| \citep{satoh2025nmfkc}. Throughout, the squared Euclidean distance is used as the objective, and the fit is reported by the coefficient of determination $R^2=\mathrm{cor}(Y,\hat{Y})^2$, the squared Pearson correlation between the observed and the fitted values over all matrix entries. We use this rather than the usual residual-based coefficient of determination $1-\sum_{p,n}(y_{p,n}-\hat{y}_{p,n})^2\big/\sum_{p,n}(y_{p,n}-\bar{y})^2$ (with $\bar{y}$ the overall mean): because the model has no intercept, this need not lie in $[0,1]$, whereas the squared correlation always does.

Two practical points, identifiability and convergence, deserve mention. The factorization $Y\approx XB$ is not unique---$X$ and $B$ are determined only up to scaling and permutation of the components---so the reported solution depends on the initialization and on the stopping rule. We initialize the columns of $X$ by $k$-means clustering of the columns of $Y$, taking the cluster centroids as the initial basis vectors; we also tried the deterministic SVD-based initialization NNDSVDar \citep{boutsidis2008}, and the fitted values and the substantive conclusions were essentially unchanged in every example. For convergence, the multiplicative updates are iterated until the relative change of the objective between successive iterations satisfies $|f^{(i)}-f^{(i-1)}|/\max(|f^{(i)}|,1)\le\varepsilon$, where $\varepsilon$ is a tolerance. We set $\varepsilon=10^{-8}$ rather than a looser $10^{-4}$: when an effect lies on the non-negativity boundary ($\theta=0$), the multiplicative updates approach it slowly, and the looser tolerance frequently stopped before convergence and split a boundary effect spuriously across components.

We demonstrate the comparison with the GCM on the orthodontic data (Section~\ref{subsec1}), the balanced design for which the model was originally proposed: there the maximum-likelihood estimator (\ref{eq16}) confirms that the NMF and GCM estimates of $\Theta$ agree, and the NMF-optimized basis fits at least as well as a prescribed straight-line design, supporting the point of Section~\ref{sec4}.

Reusing the NMF basis inside the GCM, however, is not always possible. The GCM estimator (\ref{eq16}) requires $AA'$ and the residual covariance $S$ to be invertible---which holds for the chick and Berkeley data but fails for the over-parameterized kernel covariate of the bone example ($AA'$ singular)---and the two estimates of $\Theta$ coincide only when the unconstrained estimate is essentially non-negative: they agree closely on the Berkeley data (relative difference about $2\%$) but diverge on the chick data, where some diet contrasts are negative and the non-negative NMF cannot represent them. The NMF basis is therefore not always interchangeable with a GCM design; conversely, ordinary use of the GCM requires the analyst to supply a basis $X$, hard to construct for these trajectories---the difficulty that NMF removes by optimizing $X$ from the data.

%-----------------------------------------------------------
\subsection{Growth curve data on an orthodontic measurement}\label{subsec1}
%-----------------------------------------------------------

The dataset was originally introduced in Table 1 of \citet{potthoff1964} and is available as \verb|Orthodont| in the \verb|nlme| package in R.
According to its description, \verb|Orthodont| is a dataset collected at the University of North Carolina Dental School. There were 27 children (16 males and 11 females) aged 8--14 years. Every two years, they measured the distance between the pituitary and pterygomaxillary fissures.
Thus, the observation matrix $Y$ where $P=4$ and $N=27$ consists of the \verb|distance| among 16 males and 11 females at the ages of 8, 10, 12, and 14. The values were measured using X-rays and need not increase monotonically with age.

We first perform ordinary NMF without covariates given by (\ref{eq01}), with $Q=2$ bases (equal to the number for a linear curve) and each column of the basis matrix normalized to sum to one. The fit is good: the coefficient of determination between $Y$ and $\hat{Y}=XB$ is $0.906$. The non-negative coefficients give a soft clustering directly; for an individual with coefficients $57.4$ and $53.7$ on the two bases, for instance, the normalized proportions $0.517$ and $0.483$ serve as membership probabilities.

Next, NMF is performed with the sex entering as a \emph{main-effect} covariate under reference (treatment) coding: the covariate matrix is $A=(\vec{1},\vec{m})'$, an intercept together with the male dummy $\vec{m}=(\mathrm{male}_1,\ldots,\mathrm{male}_N)'$, where $\mathrm{male}_n=1$ if individual $n$ is male and $\mathrm{male}_n=0$ if female; the second column of $\Theta$ is then the male effect on each basis---the male-minus-female difference---the quantity assessed in the inference below. (In the bone-density example of Section~\ref{subsec3}, by contrast, the same sex factor is interacted with a kernel basis and entered with indicator (one-hot) coding.)

\begin{figure}[h]
\centering
\includegraphics[width=1\linewidth]{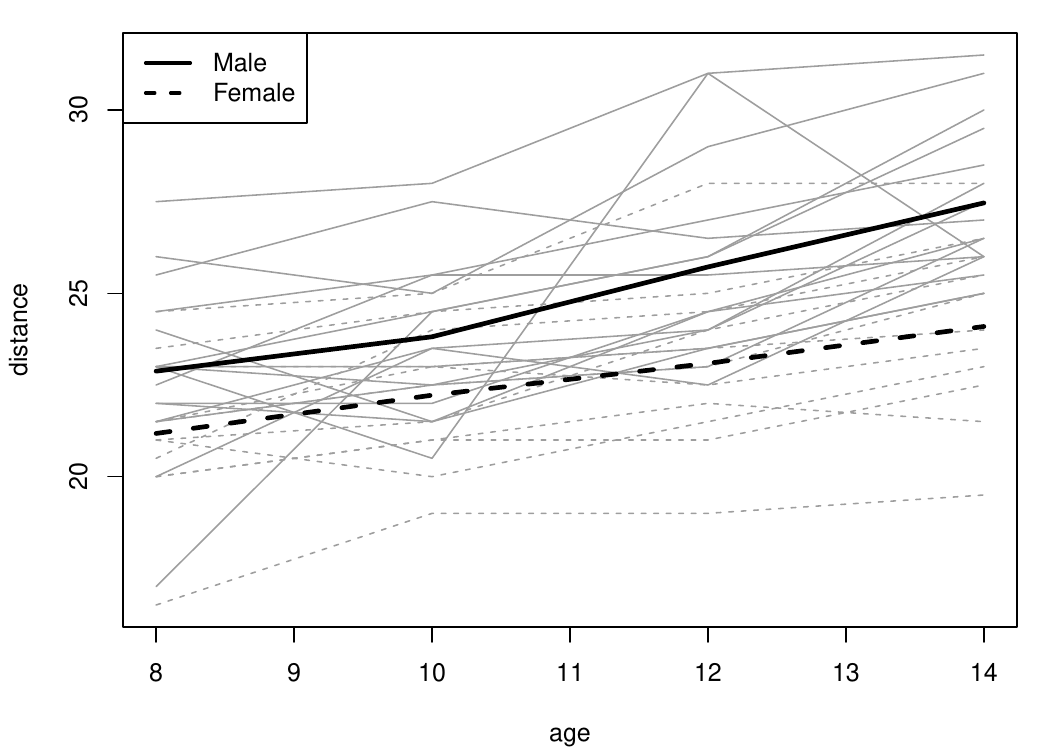}
\caption{Observation and fitted lines by NMF with covariates.
  Gray lines indicate individual observations and bold lines show the fitted curves.}\label{fig2}
\end{figure}

The optimized basis and parameter matrices and this covariate matrix are as follows, with the columns arranged so that the males come first (hence the second row of $A$ is $1$ for the males and $0$ for the females):
$$
X=
 \begin{pmatrix}
0.284&0.175\\
0.317&0.160\\
0.227&0.287\\
0.172&0.377\\
  \end{pmatrix},
\Theta=
 \begin{pmatrix}
49.036 & 0.665\\
41.550 & 8.627\\
  \end{pmatrix},
A=  
 \begin{pmatrix}
1 \cdots  1 &  1\cdots   1\\
1 \cdots  1 &  0\cdots   0\\
  \end{pmatrix}.
$$
The fitted curves are shown in Fig.~\ref{fig2}.
There was no difference in the slope of the curves for the sexes aged 8--10 years; however, after the age of 10 years, the slope for males was slightly higher.

From (\ref{eq16}), the maximum likelihood estimator, $\hat{\Theta}$, is obtained using $X$ optimized by NMF for GCM,
$$
\hat{\Theta}=
 \begin{pmatrix}
49.066 & 0.616\\
41.522 & 8.672\\
  \end{pmatrix}.
$$
The optimized $\Theta$ values obtained from the NMF and GCM estimates $\hat{\Theta}$ were almost identical.

Applying the inference of Section~\ref{sec:inference} to this fit gives Table~\ref{tab:inf_ortho}. The quantity of interest is the sex effect: it is not significant on Basis~1 (the decreasing component, $p=0.381$) but is significant on Basis~2 (the increasing component, $\hat{\theta}=8.63$, $z=3.82$, $p<0.001$), formalizing the steeper male rise noted above. The intercepts are trivially significant as baseline magnitudes and, being level parameters, should be read with the caution of Section~\ref{sec:sim}. It also reproduces the classical growth-curve-model inference: on the same $\hat{X}$, the male effect on the increasing component is significant under both the working model ($z=3.82$) and the classical GCM ($z=3.83$), and non-significant on the other under both. An ordinary straight-line GCM design $X=[\vec{1},\vec{t}]$ gives the same conclusion (male effect significant on the slope, $z=2.61$, $p<0.01$, not on the intercept).

\begin{table}[h]
\caption{Inference for the parameter matrix on the orthodontic growth data ($Q=2$). Basis~1 has loadings decreasing with age and Basis~2 increasing with age. The $p$-value is for the one-sided boundary test $H_0:\theta=0$ versus $H_1:\theta>0$.}\label{tab:inf_ortho}
\centering
\begin{tabular}{llrrrr}
\toprule
Component & Covariate & Estimate & SE & $z$ & $p$ \\
\midrule
Basis 1 (decreasing) & Intercept & 49.04 & 1.35 & 36.20 & $<0.001$ \\
                     & Male      & 0.66  & 2.20 & 0.30  & 0.381   \\
Basis 2 (increasing) & Intercept & 41.55 & 1.64 & 25.34 & $<0.001$ \\
                     & Male      & 8.63  & 2.26 & 3.82  & $<0.001$ \\
\bottomrule
\end{tabular}
\end{table}

The coefficient of determination of the covariate model based on $X$ optimized by NMF is $0.427$, which decreases to $0.422$ when $X$ is fixed to a straight-line design with the same number of bases ($Q=2$).
Therefore, even when using a growth curve model, it may be better to use $X$ optimized by NMF than $X$ given by the analyst.

%-----------------------------------------------------------
\subsection{Growth of chicks under different diets}\label{subsec:chick}
%-----------------------------------------------------------

We turn to an experiment in which the body weights of chicks on four diets were recorded at 12 time points (days 0--21), available as \verb|ChickWeight| in R; we use the 45 chicks with complete records ($P=12$, $N=45$, with $16$, $10$, $10$, and $9$ chicks on Diets~1--4). The covariate is the experimental treatment, coded by an intercept and three dummy variables (Diet~1 as the reference). A single basis already fits the data well (coefficient of determination $0.963$ without covariates and $0.776$ with the diet covariate), and a second basis barely improves it ($0.991$ and $0.781$); we nonetheless take $Q=2$ so that the non-negative coefficients give a two-component soft clustering (membership probabilities), comparable with the two-component PCA below.

\begin{figure}[h]
\centering
\includegraphics[width=0.8\linewidth]{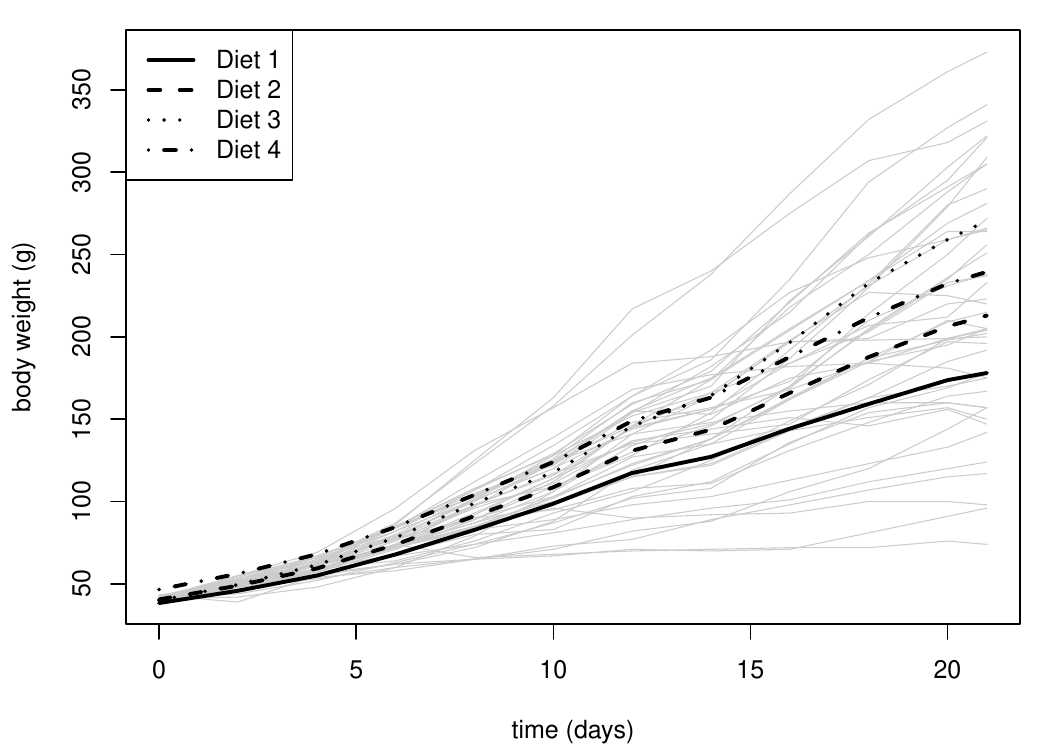}
\caption{Chick-growth data: observed trajectories (grey) and the fitted mean curves for the four diets.}\label{figchick}
\end{figure}

Table~\ref{tab:inf_chick} reports the inference. The diet effects are basis-specific: Diet~3 differs significantly from Diet~1 on the second basis ($z=3.42$, $p<0.001$) and Diet~4 on both bases ($p=0.004$ and $p=0.011$), whereas Diet~2 is not significant on either basis. Figure~\ref{figchick} shows the fitted mean curves for the four diets.

\begin{table}[h]
\caption{Inference for the parameter matrix on the chick-growth data ($Q=2$). The $p$-value is one-sided for $H_0:\theta=0$ versus $H_1:\theta>0$.}\label{tab:inf_chick}
\centering
\begin{tabular}{llrrrr}
\toprule
Component & Covariate & Estimate & SE & $z$ & $p$ \\
\midrule
Basis 1 & Intercept & 747.30 & 36.27  & 20.60 & $<0.001$ \\
        & Diet 2    & 39.59  & 52.97  & 0.75  & 0.227   \\
        & Diet 3    & 21.00  & 51.75  & 0.41  & 0.342   \\
        & Diet 4    & 158.08 & 58.88  & 2.68  & 0.004   \\
Basis 2 & Intercept & 541.12 & 67.41  & 8.03  & $<0.001$ \\
        & Diet 2    & 142.16 & 123.68 & 1.15  & 0.125   \\
        & Diet 3    & 405.08 & 118.37 & 3.42  & $<0.001$ \\
        & Diet 4    & 220.33 & 96.81  & 2.28  & 0.011   \\
\bottomrule
\end{tabular}
\end{table}

\begin{figure}[h]
\centering
\includegraphics[width=0.95\linewidth]{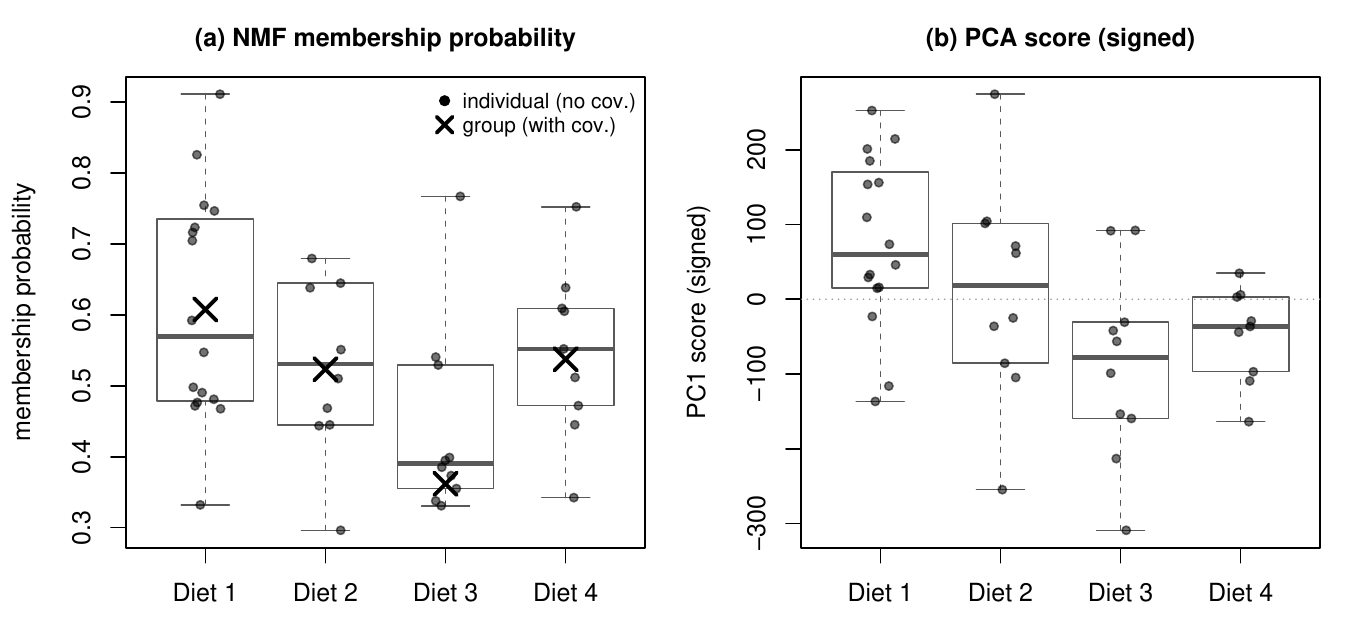}
\caption{Chick-growth data, by diet (box plots overlaid on jittered points): (a) the non-negative NMF coefficients normalize to membership probabilities in $[0,1]$, with the cross ($\times$) marking the per-group membership from the covariate model ($B=\Theta A$), which sits at the centre of the individual distribution; (b) the first principal component score (the projection component most associated with diet) is signed. Both vary with diet, but only NMF provides a soft-clustering membership.}\label{figchickpca}
\end{figure}

\paragraph{Comparison with principal component analysis.}
Because the diets affect growth, the non-negative coefficients separate the groups as membership probabilities, which makes this example well suited to a comparison with principal component analysis (PCA). A two-component PCA reconstructs slightly better (coefficient of determination $0.997$ versus $0.991$), but the essential difference is interpretability: the NMF coefficients normalize to membership probabilities in $[0,1]$ (here ranging over $[0.30,0.99]$), whereas the PCA scores are signed and require a downstream clustering step to recover groups. For the projection method we display the component most associated with the grouping variable; here this is the first principal component (which also dominates the variation), whose scores range over $[-309,274]$. Figure~\ref{figchickpca} contrasts the two by diet, with box plots overlaid on the jittered points to make the group distributions visible. In particular, Diet~3 stands out in Figure~\ref{figchickpca}(a): its membership probabilities are the lowest, forming a cluster separate from the other diets, consistent with Figure~\ref{figchick}, where the Diet~3 mean curve attains the highest weights.

%-----------------------------------------------------------
\subsection{Growth in stature (Berkeley growth study)}\label{subsec:berkeley}
%-----------------------------------------------------------

We next analyse the heights of 93 children (39 boys and 54 girls) measured at 31 ages between 1 and 18 years, available as \verb|growth| in the \verb|fda| package; the observation matrix $Y$ has $P=31$ and $N=93$. Using the intercept and a male indicator as covariates, we take $Q=2$ bases; as in the chick example a single basis fits comparably, but two bases let the sex effect be resolved by component (and match the two-harmonic FPCA below). With $Q=2$ the coefficient of determination is $0.997$ without covariates and $0.968$ with the sex covariate. As in the orthodontic example, the male effect is significant on only one basis---the component whose loadings increase with age ($\hat{\theta}=128.0$, $z=10.7$, $p<0.001$)---and not on the other ($p=0.5$); see Table~\ref{tab:inf_berkeley}. The two growth data sets thus give a consistent finding: sex acts selectively on the later-growth component.

\begin{table}[h]
\caption{Inference for the parameter matrix on the Berkeley growth data ($Q=2$). The $p$-value is one-sided for $H_0:\theta=0$ versus $H_1:\theta>0$.}\label{tab:inf_berkeley}
\centering
\begin{tabular}{llrrrr}
\toprule
Component & Covariate & Estimate & SE & $z$ & $p$ \\
\midrule
Basis 1 & Intercept & 2765.95 & 16.44 & 168.2 & $<0.001$ \\
        & Male      & 0.00    & 24.44 & 0.00  & 0.500   \\
Basis 2 & Intercept & 1422.84 & 7.50  & 189.8 & $<0.001$ \\
        & Male      & 128.04  & 11.99 & 10.68 & $<0.001$ \\
\bottomrule
\end{tabular}
\end{table}

\begin{figure}[h]
\centering
\includegraphics[width=0.8\linewidth]{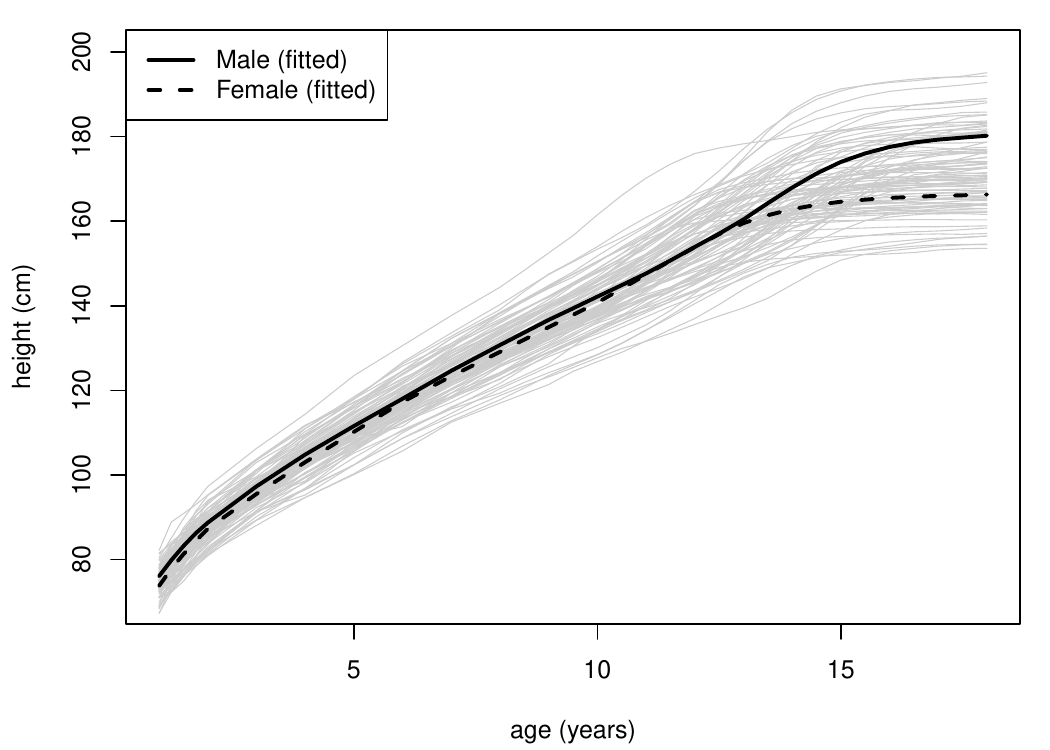}
\caption{Berkeley growth data: observed trajectories (grey) and the fitted mean curves for boys and girls.}\label{figberkeley}
\end{figure}

Figure~\ref{figberkeley} shows the fitted mean curves for the two sexes: they coincide in childhood and diverge only after puberty---the late divergence that a single basis, which can express the sex difference only as a uniform scaling of one common shape, cannot capture.

\begin{figure}[h]
\centering
\includegraphics[width=0.95\linewidth]{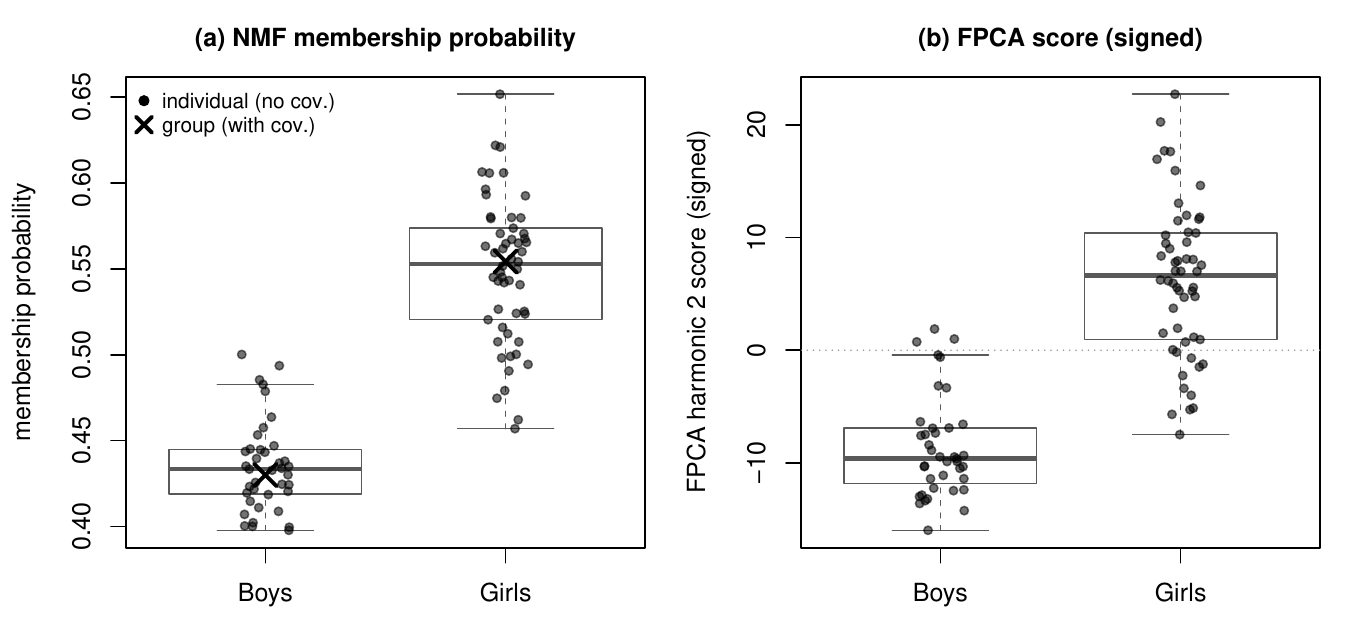}
\caption{Berkeley growth data, by sex (box plots overlaid on jittered points): (a) the non-negative NMF coefficients normalize to membership probabilities in $[0,1]$, with the cross ($\times$) marking the per-group membership from the covariate model ($B=\Theta A$), which sits at the centre of the individual distribution; (b) the second FPCA harmonic score (the harmonic most associated with sex; the first captures overall size) is signed. Both separate the sexes, but only NMF provides a soft-clustering membership.}\label{figberkeleyfpca}
\end{figure}

\paragraph{Comparison with functional PCA.}
Because these are smooth growth curves, the natural projection-based competitor is functional principal component analysis (FPCA). Its first two harmonics explain $95.5\%$ of the curve variation, comparable to the NMF reconstruction. The first harmonic captures overall body size and is only weakly related to sex; as for the chick example we display the harmonic most associated with the grouping variable, which here is the second harmonic, whose signed scores range over $[-16.0,\,22.7]$ and, like ordinary PCA, require a downstream classifier to produce groups. The non-negative NMF coefficients, by contrast, are membership probabilities in $[0,1]$. This illustrates a structural difference between the two decompositions: the FPCA harmonics are ordered by the variance they explain, so the leading harmonic need not be the one that discriminates the groups (here it is the second), whereas the NMF bases carry no such ordering and are interchangeable, each contributing a non-negative membership component on an equal footing. Figure~\ref{figberkeleyfpca} contrasts the two by sex, with box plots overlaid on the jittered points; both relate to sex (correlations $0.84$ and $-0.78$ with the male indicator), but only NMF expresses it as a probability. The covariate model supplies the same membership directly: its coefficients $B=\Theta A$ give one probability per sex, marked by crosses in Fig.~\ref{figberkeleyfpca}(a), that separate the sexes and lie at the centre of the covariate-free individual memberships---the soft-clustering counterpart of the significant sex effect in Table~\ref{tab:inf_berkeley}.

%-----------------------------------------------------------
\subsection{Spinal bone mineral density across adolescence}\label{subsec3}
%-----------------------------------------------------------

As a final example we combine the Gaussian kernel covariate with the multiple-regression special case of Section~\ref{sec4}. The \verb|bone| data of \citet{hastie2009}, originally collected by \citet{bachrach1999}, record the relative change in spinal bone mineral density (BMD) for $485$ measurements on adolescents ($259$ on females, $226$ on males) as a function of age. Each measurement is a scalar, so the observation matrix has a single row ($P=1$): writing $\vec{y}'=(y_1,\ldots,y_N)$ with $N=485$, this is the multiple-regression special case in which $X=1$ and $\vec{y}'\approx\vec{\theta}'A$ derived in Section~\ref{sec4} (a small constant shift makes the few negative relative changes non-negative).

We fit the varying-coefficient model $y_i=f_0(\mathrm{age}_i)+\mathrm{male}_i\,f_1(\mathrm{age}_i)$ \citep{hastie1993} with the female curve as control, so $f_0$ is the female curve and $f_1=f_M-f_F$ the male effect. Each sex curve is expanded in a Gaussian kernel of age, $k(u,u')=\exp(-\beta|u-u'|^{2})$ \citep{satoh2023}. Rather than one kernel per observation, we centre kernels at $M=18$ landmark ages on a one-year grid spanning the data (a Nystr\"om low-rank construction, \citealp{williams2000}), giving the $M\times N$ matrix $(K)_{m,n}=k(\vec{u}^{*}_m,\vec{u}_n)$. The two non-negative curves are $f_F=\vec{\theta}_F'K$ (females) and $f_M=\vec{\theta}_M'K$ (males). Coding sex one-hot as $\vec{d}_n=(1-\mathrm{male}_n,\ \mathrm{male}_n)'$ and writing $\vec{k}_n$ for the $n$th column of $K$, the covariate matrix stacks the sex-by-kernel interaction column by column,
\begin{equation*}
A=(\vec{d}_1\otimes\vec{k}_1,\ \ldots,\ \vec{d}_N\otimes\vec{k}_N)\in\mathbb{R}^{2M\times N},\qquad
\vec{\theta}=\begin{pmatrix}\vec{\theta}_F\\ \vec{\theta}_M\end{pmatrix}\in\mathbb{R}^{2M},\ \vec{\theta}\ge \vec{0},
\end{equation*}
where $\otimes$ is the Kronecker product, $\vec{y}'$ is $1\times N$, and $M=18$, $N=485$. This is the multiple-regression special case $\vec{y}'\approx\vec{\theta}'A$ of Section~\ref{sec4}, fitted by the multiplicative updates of Section~\ref{sec3}. The one-hot coding---rather than a treatment-coded male dummy carrying a single non-negative coefficient---leaves the male effect $f_1=f_M-f_F$ free to change sign, so the male curve may fall below or rise above the female one. The scale $\beta$ was selected by cross-validation over a seven-point grid centred on the median-heuristic value, giving $\beta\approx0.11$ and $R^2=0.43$.

\begin{figure}[h]
\centering
\includegraphics[width=0.8\linewidth]{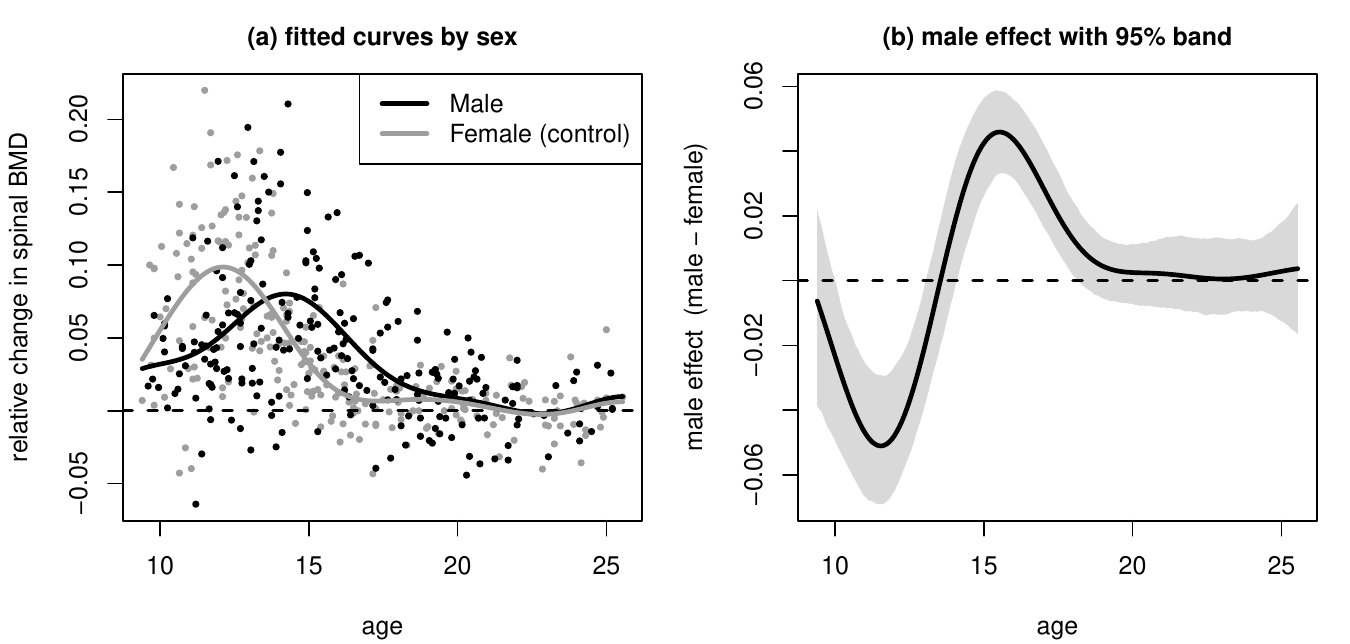}
\caption{Spinal bone mineral density (BMD), with the female curve taken as the control. (a) Relative change in BMD versus age: points are individual measurements (males black, females gray), and the curves are the fitted female ($f_0=f_F$) and male ($f_0+f_1=f_M$) trajectories. (b) The estimated male effect $f_1=f_M-f_F$ (male minus female) versus age, with a pointwise $95\%$ confidence band from the wild bootstrap.}\label{figbone}
\end{figure}

Figure~\ref{figbone}(a) shows the fitted curves: female BMD accrual peaks earlier (around age~11--12) and declines sooner, while the male curve peaks later (around age~14). Panel~(b) shows the male effect $f_1=f_M-f_F$ with a pointwise $95\%$ confidence band. The individual landmark coefficients are not separately identifiable (the kernel covariate is over-parameterized, so $AA'$ is rank-deficient), but the contrast $f_1(\mathrm{age})$ is the kind of quantity the simulation of Section~\ref{sec:sim} found to be well calibrated. As the closed-form covariance is unavailable, the band is obtained by the re-estimation wild bootstrap of Section~\ref{sec:inference} ($B=1000$, Rademacher multipliers; cf.\ \citealt{mammen1993} for an asymmetric alternative): in each replication the model is re-fitted and $f_1$ recomputed, and the band is the pointwise $2.5\%$--$97.5\%$ quantiles. The effect is significantly negative over roughly ages~$10$--$13$ (males lagging), crosses zero near age~$13$--$14$, and is significantly positive over ages~$14$--$18$ (males then exceeding females). The sex effect is thus a \emph{function} of age rather than the single number of the orthodontic and Berkeley examples, reproducing the well-known sex difference in the timing of the adolescent growth spurt.

%-----------------------------------------------------------
\section{Simulation study}\label{sec:sim}
%-----------------------------------------------------------

We validate the proposed inference of Section~\ref{sec:inference} by a simulation study calibrated to the orthodontic growth data of Section~\ref{subsec1}. Taking the estimated structure there (Table~\ref{tab:inf_ortho}) as the ground truth, we generate data from the working model (\ref{eq:workmodel}) with $\hat{X}$ and $\sigma^2$ calibrated to that example ($P=4$, $Q=2$, $\sigma^2=5.06$), using the intercept and the male indicator as covariates. The true parameters, rounded from Table~\ref{tab:inf_ortho}, include both interior and boundary values: on the increasing-loading component, intercept $=41$ and male $=8$; on the other, intercept $=49$ and male $=0$ (the latter giving an exact boundary null). We replicated the design at sample sizes $N=27,108,270$ (that is, $N=27k$) with $2000$ replications each. In every replication the data are generated from the same true basis; the two modes differ only in the basis that is supplied to the inference: (i) the true basis is used (the basis is treated as known, isolating the inference for $\Theta$); and (ii) the basis is re-estimated by NMF from each simulated data set and the inference is conditioned on this estimate $\hat{X}$, mimicking the practical situation in which the basis is unknown. Table~\ref{tab:inf_sim} reports the bias and the empirical coverage of nominal $95\%$ Wald intervals.

When the basis is fixed (mode i), the coverage is at or slightly above the nominal level for all parameters---the standard errors are mildly conservative---the bias vanishes as $N$ grows, and the one-sided test controls its size: for the boundary parameter (true male effect $=0$) the empirical type-I error was $0.04$--$0.06$, while the power for the true effect (male $=8$) rose from $0.80$ at $N=27$ to $1.00$ for $N\ge 108$ (type-I error and power are not shown in Table~\ref{tab:inf_sim}).

When the basis is re-estimated (mode ii), a clear dichotomy appears. Inference for the \emph{covariate-effect contrast} (the male effect) remains well calibrated, with coverage close to nominal ($0.90$--$0.99$), because a contrast between groups is insensitive to the scale and rotation of the basis. Inference for the \emph{overall level} (the intercepts), in contrast, under-covers substantially ($0.67$--$0.79$) and does not improve with $N$, because the level is confounded with the estimated scale of the basis, an uncertainty that the conditional standard errors do not capture. We therefore recommend the proposed inference for assessing covariate effects, the quantities of primary scientific interest, while the intercepts should be interpreted with caution.

\begin{table}[h]
\caption{Simulation results: bias and empirical coverage of nominal $95\%$ Wald confidence intervals, with the basis fixed at its true value and with the basis re-estimated in each replication ($2000$ replications). Subscripts index the basis component; the true value of each parameter is given in parentheses.}\label{tab:inf_sim}
\centering
\begin{tabular}{lrrrrr}
\toprule
 & & \multicolumn{2}{c}{Basis fixed} & \multicolumn{2}{c}{Basis re-estimated} \\
\cmidrule(lr){3-4}\cmidrule(lr){5-6}
Parameter & $N$ & bias & cov.\ & bias & cov.\ \\
\midrule
Intercept$_1$ ($49$) & 27  & $-0.73$ & 0.979 & $0.59$  & 0.685 \\
                     & 108 & $-0.41$ & 0.967 & $-0.36$ & 0.791 \\
                     & 270 & $-0.27$ & 0.974 & $-0.64$ & 0.713 \\
Intercept$_2$ ($41$) & 27  & $0.55$  & 0.974 & $-0.69$ & 0.668 \\
                     & 108 & $0.32$  & 0.971 & $0.31$  & 0.792 \\
                     & 270 & $0.20$  & 0.973 & $0.60$  & 0.722 \\
Male$_1$ ($0$)       & 27  & $1.28$  & 0.978 & $1.14$  & 0.922 \\
                     & 108 & $0.71$  & 0.968 & $0.81$  & 0.989 \\
                     & 270 & $0.48$  & 0.968 & $0.82$  & 0.967 \\
Male$_2$ ($8$)       & 27  & $-1.00$ & 0.982 & $-1.01$ & 0.898 \\
                     & 108 & $-0.57$ & 0.976 & $-0.78$ & 0.974 \\
                     & 270 & $-0.38$ & 0.975 & $-0.82$ & 0.943 \\
\bottomrule
\end{tabular}
\end{table}

%-----------------------------------------------------------
\section{Discussion and Conclusion}\label{sec6}
%-----------------------------------------------------------

We applied NMF with covariates to longitudinal growth data and made explicit the formal equivalence, in mean structure, between NMF with covariates and the growth curve model (GCM). The comparison showed that the NMF-optimized basis can serve as a GCM design---the two giving consistent estimates of $\Theta$ when the covariate effects are non-negative (the orthodontic and Berkeley data) and diverging otherwise (some chick diets)---so NMF supplies the data-driven basis the GCM would otherwise require. The main methodological contribution is statistical inference for the parameter matrix $\Theta$: treating the NMF-optimized basis as a known design and adopting a Gaussian working model, we obtained standard errors, Wald-type tests, and confidence intervals for the covariate effects, with one-sided boundary tests reflecting the non-negativity constraint. This extends to the data-driven-basis setting of NMF the inference that the GCM offers when the basis is known.

The simulation study of Section~\ref{sec:sim} delineated the scope of this inference. With the basis treated as known it is well calibrated and the one-sided test controls its size. When the basis is re-estimated, inference on covariate-effect \emph{contrasts}---the quantities of primary scientific interest---remains well calibrated, because contrasts are insensitive to the scale and rotation of the estimated basis, whereas inference on the overall \emph{level} is anti-conservative, the level being confounded with the estimated scale of the basis. We therefore recommend the proposed inference for covariate effects and advise caution for level parameters. The real-data examples were consistent with this: on the orthodontic and Berkeley growth data the sex effect was significant only on the age-increasing component, and on the orthodontic data the conditional inference reproduced the classical growth-curve-model inference.

Practically, because the multiplicative updates involve no matrix inversion, the optimization stays numerically stable even when the observed matrix is large and the parameters are many---the very property that made NMF attractive for high-dimensional image analysis such as the parts-based representation of facial images and face recognition \citep{lee1999,chen2022}. Gaussian-kernel covariates, with the scale chosen by cross-validation, accommodate flexible covariate structures while retaining predictive ability; the bone-density example illustrated this as a non-negative varying-coefficient model and as the multiple-regression special case of the framework, with a pointwise confidence band locating the ages at which the sexes' bone accrual differs. Because the non-negative coefficients normalize to membership probabilities, NMF with covariates combines aspects of regression and of soft clustering---unlike the signed scores of PCA and FPCA, which need a downstream classifier, these coefficients cluster directly---making it useful both for exploratory analysis and for prediction from covariates; the NMF with covariates is implemented in the \verb|nmfkc| R package. A limitation is that the inference is conditional on the optimized basis: fully propagating the uncertainty of basis estimation, together with the choice of the rank $Q$, is left for future work. More broadly, the present study is one instance of using NMF with covariates to connect NMF with established models---here the growth curve model---and companion works develop the connections to vector autoregression for multivariate time series \citep{satoh2026var} and to classification through a label matrix \citep{satoh2026lab}.

\bmhead{Funding}

This work was partly supported by JSPS KAKENHI Grant Numbers 22K11930, 24K03007, 25H00482, 25K15229, the project research fund by the Fuji Seal Foundation.

\bmhead{Disclosure of interest}

The author has no competing financial or non-financial interests to disclose.

\bmhead{Data availability statement}

All data sets analysed in this study are publicly available in R: \verb|Orthodont| (the \verb|nlme| package), \verb|growth| (the \verb|fda| package), \verb|ChickWeight| (base R), and \verb|bone| (the \verb|ElemStatLearn| package; \citet{bachrach1999,hastie2009}).

\bmhead{Code availability}

The NMF with covariates in this paper was computed using the R package \verb|nmfkc| \citep{satoh2025nmfkc}, publicly available on CRAN at \url{https://CRAN.R-project.org/package=nmfkc} (\url{https://doi.org/10.32614/CRAN.package.nmfkc}).

\bmhead{Generative AI disclosure}

The author used Anthropic's Claude (accessed through Claude Code; Claude Opus and Claude Sonnet 4-series models, 2025--2026) as an assistant in the revision of R scripts and for English-language editing and restructuring of this manuscript. All methodological choices, the theoretical results, the numerical experiments, and the data analyses were designed, verified, and approved by the author, who takes full responsibility for the content of the article. No research data, results, figures, or tables were generated by artificial intelligence.

\bmhead{Author contributions}

The author confirms sole responsibility for the study conception and design, the methodology and analysis, and the preparation of the manuscript.

\bmhead{ORCID}

Kenichi Satoh
https://orcid.org/0000-0003-4436-9347

\bibliography{nmfasgcm}% common bib file
\end{document}